\documentclass[conference]{IEEEtran}
\IEEEoverridecommandlockouts
\usepackage{amsmath,amssymb,amsfonts}
\usepackage{graphicx}
\usepackage{textcomp}
\usepackage{xcolor}
\def\BibTeX{{\rm B\kern-.05em{\sc i\kern-.025em b}\kern-.08em
    T\kern-.1667em\lower.7ex\hbox{E}\kern-.125emX}}
\usepackage{algpseudocode,algorithm}
\usepackage{booktabs}
\usepackage{hyperref}
\usepackage[backend=biber,style=ieee,natbib=true,doi=true]{biblatex} 
\usepackage{tikz}
\usepackage{textcomp}
\usepackage{hyperref}
\usepackage{lipsum}

\newcommand\copyrighttext{%
  \footnotesize \textcopyright 2022 IEEE. Personal use of this material is permitted.
  Permission from IEEE must be obtained for all other uses, in any current or future 
  media, including reprinting/republishing this material for advertising or promotional 
  purposes, creating new collective works, for resale or redistribution to servers or 
  lists, or reuse of any copyrighted component of this work in other works. 
  DOI: \href{http://doi.org/10.1109/CSR54599.2022.9850338}{10.1109/CSR54599.2022.9850338}}
\newcommand\copyrightnotice{%
\begin{tikzpicture}[remember picture,overlay]
\node[anchor=south,yshift=10pt] at (current page.south) {\fbox{\parbox{\dimexpr\textwidth-\fboxsep-\fboxrule\relax}{\copyrighttext}}};
\end{tikzpicture}%
}

\begin{document}

\copyrightnotice

\title{A Bayesian Model Combination based approach to Active Malware Analysis\\
\thanks{This research is supported by the European Union’s Horizon 2020 Research and Innovation program CONCORDIA under Grant Agreement No. 830927}
}

\author{\IEEEauthorblockN{1\textsuperscript{st} Abhilash Hota}
\IEEEauthorblockA{\textit{Computer Science} \\
\textit{Jacobs University}\\
Bremen, Germany \\
a.hota@jacobs-university.de}
\and
\IEEEauthorblockN{2\textsuperscript{nd} Jürgen Schönwälder}
\IEEEauthorblockA{\textit{Computer Science} \\
\textit{Jacobs University}\\
Bremen, Germany \\
j.schoenwaelder@jacobs-university.de}
}


\maketitle

\begin{abstract}
    Active Malware Analysis involves modeling malware behavior by executing actions to trigger responses and explore multiple execution paths. One of the aims is making the action selection more efficient. This paper treats Active Malware Analysis as a Bayes-Active Markov Decision Process and uses a Bayesian Model Combination approach to train an analyzer agent. We show an improvement in performance against other Bayesian and stochastic approaches to Active Malware Analysis.
\end{abstract}


\section{Introduction}

Mobile devices have become increasingly ubiquitous and used in various critical functions such as banking. The Android OS has a huge share of the smartphone market \cite{statcounter} and as such, has become a preferred target for malware authors in recent years. The number of new malwares has been steadily increasing, with the AV-Test institute registering an average of around 250000 new samples for analysis every month last year \cite{avtest}. Users of Android OS devices have access to multiple marketplaces, aside from the official Google Play store, from where they can download and install applications. Users can also download and install applications not currently hosted on the marketplaces. The process of malware analysis generally tends to involve largely manual work by analysts. However, analyzing the large number of new applications being released on various Android application marketplaces, forces malware analysts to rely on various techniques to automate the malware analysis process.

Malware analysts try to model a malware's behavior on execution and use this information to develop signatures that can then be used for malware detection on a network or device. Approaches to malware analysis can be broadly classified as static or dynamic. In static analysis \cite{wichmann1995industrial} an application is examined without executing the actual instructions. This approach is however vulnerable to various evasion techniques employed by malware authors like packing or code obfuscation\cite{moser2007limits}. Dynamic analysis, on the other hand, involves executing an application in a sandbox environment and recording the execution traces. Traditionally, dynamic analysis has been passive in that an application is simply executed and its behavior observed. A lot of application behaviors however, especially with mobile applications, are often only triggered by user interaction.

Active Malware Analysis (AMA) involves modeling a malware sample's behavior by executing actions like button clicks, sending messages and microphone activation in order to trigger responses and explore multiple execution paths. A major focus in this approach is how to efficiently trigger different behaviors from an application that might require user interaction in order to explore the execution paths. Initial efforts in AMA have been based either on past recorded user activity \cite{10.1007/978-3-319-11203-9_11}, or a pseudo-random selection of possible triggering actions \cite{BHANDARI201846,MARTIN2018121}. More recent work has focused on employing machine learning techniques to make smarter choices about which actions to use to trigger these responses. 

Our work models AMA as a Bayes-Adaptive Markov Decision Process (BAMDP) \cite{martin1967bayesian} and applies Bayesian Model Combination (BMC) \cite{} and $\epsilon$-BMC \cite{} to the problem. The intent here is to explore the use of model-free reinforcement learning using the epsilon-greedy exploration policy to resolve the exploration-exploitation trade-off in AMA. We evaluate the approach by training the model on a dataset of 15000 Android malware samples and provide a comparison against BAMA\cite{10.5555/3398761.3398901} and Bayes-adaptive Monte Carlo Planning \cite{} as approaches to AMA. Comparison is also provided against a stochastic game based approach using Monte Carlo Tree Search models that have been trained on the same dataset.

The rest of the paper is structured as follows: Section \ref{sec:background} provides some background information and discusses related work. Section \ref{sec:model} the model designs implemented in this paper. Section \ref{sec:experiment} explains the experimental setup and the process followed by the agent for malware analysis. Section \ref{sec:evaluation} provides an evaluation of the results and we conclude the paper in Section \ref{sec:conclusions}.

\section{Background and Related Work}
\label{sec:background}

In this section we provide the relevant background information on Markov decision processes, Bayesian model combination and AMA.

\subsection{Markov Decision Process}

The Markov Decision Process (MDP) is an approach to sequential decision-making in Markovian dynamical systems \cite{bellman1957markovian}. It tracks a system  state that can change according to performed actions and affects the outcome of the system.

An MDP M is a tuple $(S, A, P, P_0, q)$ where
\begin{itemize}
    \item S is the set of states,
    \item A is the set of actions,
    \item $P(\cdot | s,a) \in P(S)$ is the probability distribution over next states, when action $a$ is taken in state $s$,
    \item $P_0 \in P(S)$ is the probability distribution according to which the initial state is selected, and
    \item $R(s,a) \sim q(\cdot | s,a) \in P(R)$ is a random variable representing the reward obtained when action $a$ is taken in state $s$.
\end{itemize}

The MDP controlled by a policy $\mu$ induces a Markov chain $M^\mu$ with reward distribution $q^{\mu}(\cdot|s) = q(\cdot|s, \mu(s))$ such that $R^{\mu}(s) = R(s, \mu(s)) \in q^{\mu}(·|s)$, transition kernel $P^{\mu}(\cdot|s) = P ( \cdot|s, \mu(s))$, and stationary distribution over states $\pi^\mu$. A discount factor $\gamma \in [0,1]$ determines the exponential devaluation rate of delayed rewards.

The discounted return of a state $s$ is defined as the sum of discounted rewards that the agent encounters when starting in state $s$ and following some policy $\mu$ afterwards.

\begin{multline}\label{eq1}
    D^{\mu}(s) = \sum_{t=0}^{\infty}\gamma^{t}R(Z_{t})|Z_0 = (s,\mu(\cdot|s)), \\
    with  \;S_{t+1} \sim P^{\mu}(\cdot|S_{t})
\end{multline}

The expected value of $D^\mu$ gives us the value function of the policy. The value of a state $s$ under some policy $\mu$ can be represented in terms of its immediate reward and the values of its successor states using the Bellman equation as

\begin{equation}\label{eq2}
    V^{\mu}(s) = R^{\mu}(s) + \gamma\int_{S}P^{\mu}(s^\prime|s)V^{\mu}(s^\prime)ds^\prime
\end{equation}

\subsection{Reinforcement Learning}

Reinforcement learning (RL) is a field of machine learning that aims at learning suitable actions to maximize the reward in a particular situation. In RL the transition and reward functions are unknown. Hence optimal policies are learned from experience. These are defined as sequences of transitions $(s_t, a_t, r_{t+1}, s_{t+1}, a_{t+1}), t = 0, 1, \cdots$ broken up into episodes. States and rewards are sampled from the environment, and actions follow some exploration policy $\pi$.

Given an estimate $G_t$ of the expected return at time $t$ starting from state $s$ and taking action $a$, temporal difference learning updates the expected return as follows:
\begin{equation}
    Q_{t+1}(s, a) = Q_{t}(s, a) + \eta _{t}(G_t - Q_{t}(s, a))
\end{equation}

where $\eta _{t} \in (0, 1]$ is a learning rate parameter and varies by the application domain. $G_t$t is generally bootstrapped from the current Q-values. Popular bootstrapping algorithms include Q learning \cite{} and Expected State–action–reward–state–action (Expected SARSA) \cite{}.

Q learning \cite{} is an off-policy algorithm. It evaluates and improves a different policy than the one being used to select actions. Under certain conditions Q learning has been shown to converge to the optimal policy with a probability of 1 \cite{}. $G_t$ is bootstrapped using Q learning as follows:
\begin{equation}
    G_{t}^{Q} = r_{t+1} + \gamma\max_{a' \in A}Q_{t}(s_{t+1}, a')
\end{equation}
where $\gamma \in (0,1)$ is a discount factor.

Expected SARSA \cite{} is an on-policy algorithm. It evaluates and improves the policy currently being used for action selection. In expected SARSA the uncertainty of the next action $a_{t+1}$ is averaged out with respect to the policy $\pi$, resulting in a significant reduction in variance \cite{}. $G_t$ using expected SARSA is given by
\begin{equation}
    G_{t}^{ExpSARSA} = r_{t+1} + \gamma E_{a' \sim \pi}[Q_{t}(s_{t+1}, a')]
\end{equation}

\subsection{Bayes-Adaptive Markov Decision Process}

The Bayes-Adaptive Markov Decision Process (BAMDP) is an extension of the conventional MDP model. The state space of the BAMDP combines the initial set of states $S$, with the posterior parameters on the transition function. This joint space is called a hyper-state. Transitions between hyper-states are captured in the BAMDP transition model.

A BAMDP $M$ is a tuple $(S^\prime, A^\prime, P^\prime, P^\prime_0, R^\prime)$ where
\begin{itemize}
    \item $S^\prime$ is the set of hyper-states,
    \item $A^\prime$ is the set of actions,
    \item $P^\prime(\cdot|s, \phi, a)$ is the transition function between hyper-states, when action $a$ is taken in hyper-state $(s, \phi)$,
    \item $P^\prime_0 \in P(S × \phi)$ combines the initial distribution over states with the prior over transition functions, and
    \item $R^\prime(s, \phi, a) = R(s, a)$ represents the reward obtained when action $a$ is taken in state $s$.
\end{itemize}

The value function of the BAMDP can be expressed using the Bellman equation:

\begin{multline}\label{eq3}
    V^*(s,\phi) = \max_{a \in A}[ R^\prime(s, \phi, a) + \gamma \sum_{(s^\prime,\phi^\prime)}{P^\prime(s^\prime,\phi^\prime|a, \phi, a)V^*(s^\prime,\phi^\prime)} ]\\
    =  \max_{a \in A}[ R(s,a) + \gamma\sum_{s^\prime \in S}{\frac{\phi^{a}_{s, s^\prime}}{s^n \in S^{\phi^{a}_{s,s^{\prime\prime}}}}V^*(s^\prime,\phi^\prime)}]
\end{multline}

The posterior over the transition function is represented by a Dirichlet distribution which is formulated based on prior knowledge of the application domain.

\subsection{Exploration vs Exploitation}

A major issue in reinforcement learning is exploration, i.e., how the agent should choose actions while learning about a task. Exploitation, on the other hand, looks at which actions are to be selected in order to maximize the expected reward with respect to the current value function estimate. Exploration here uses epsilon greedy policies defined as

\begin{equation}
    \pi_{t}(s,a) = \begin{cases}
                 1 - \epsilon + \frac{\epsilon}{|A|} & \text{if a = $argmax_{a^{\prime}Q_{t}(s,a^\prime)}$}\\
                 \frac{\epsilon}{|A|} & \text{otherwise}
\end{cases}
\end{equation}

Either a random action is chosen from the action space $A$ for exploration, with some probability $\epsilon_{t} \in [0, 1]$, or a greedy action is selected for exploitation according to $Q_t$. The optimal value for $\epsilon$ is generally problem-specific, and determined empirically. $\epsilon$ is typically annealed in order to favor exploration in the beginning, and exploitation closer to convergence \cite{}[Sutton and Barto, 2018].

\subsection{Bayesian Learning}

Bayesian learning involves making inferences like point and interval estimates about some random variable $X$ from a probability distribution over $X$. The process of inferring $X$ from samples of some random variable $Y$ involves the following general steps:
\begin{enumerate}
    \item Choose a probability density $P(X)$, called the prior distribution, that represents beliefs about the random variable $X$ based on prior domain knowledge
    \item Select a statistical model $P(Y|X)$ that represents a statistical dependence between $X$ and $Y$
    \item Gather data points for $Y$
    \item Update beliefs about $X$ by computing the posterior distribution using Bayes rule
    \begin{equation}\label{eq5}
        P(X|Y = y) = \frac{P(y|X)P(X)}{\int P(y|X^\prime)P(X^\prime)dX^\prime}
    \end{equation}
    
\end{enumerate}

\subsection{Bayesian Model Combination}

\begin{algorithm}
    \begin{algorithmic}[1]
        \caption{$\epsilon$ - Bayesian Model Combination}
        \label{alg:epsilonBMC}
        \Procedure{$\epsilon$-BMC}{$\mu _0, \tau _0, a_0, b_0, \hat{\mu} = 0, \hat{\sigma}^2 = \infty, \infty, \beta$}
            Initialize $s$ at root
            \For{\texttt{each episode}}
                \For{\texttt{each step in the episode}}
                    \State{$\epsilon \gets \frac{\alpha}{\alpha + \beta} $}
                    \State{choose action a using $\epsilon$-greedy policy $\pi$}
                    \State{take action a, observe r and s'}
                    \State{$G^Q \gets r + \gamma \max_{a'}Q(s', a')$}
                    \State{$G^U \gets r + \gamma \frac{1}{|A|}\sum_{a'}Q(s', a')$}
                    \State{$G^ExpS \gets r + \gamma \sum_{a'} \pi^{\epsilon}(a'|s')Q(s', a')$}
                    \State{$Q(s,a) \gets Q(s,a) + \eta [G^ExpS - Q(s,a)]$}
                    \State{Update $\hat{\mu}$ and $\hat{\sigma}^2$ using $G^ExpS$}
                    \State{Compute a, b, $e^Q$ and $e^U$}
                    \State{Update $\alpha$ and $\beta$}
                    \State{$s \gets s'$}
                \EndFor
            \EndFor
        \EndProcedure
    \end{algorithmic}
\end{algorithm}

\subsection{Active Malware Analysis}

AMA aims to develop dynamic analysis systems that perform actions in order to trigger different behaviors of the malware being analyzed. Payload deployment can often be hidden behind conditional requirements that look for specific user actions, especially in mobile malware. Initial approaches to AMA have selected user inputs based on existing data about user behavior patterns or simply used a pseudo-random selection of possible actions. 

Suarez-Tangil et al.~\cite{10.1007/978-3-319-11203-9_11} propose an analyzer that uses stochastic models extracted from samples of real recorded user behaviors to reproduce specific conditions in order to trigger malicious behavior. Bhandari et al. \cite{BHANDARI201846} propose using random triggers in a runtime semantic-aware malware detector that is hardened against code injection. CANDYMAN \cite{MARTIN2018121} introduces a Markov chain model that expresses the malicious dynamics of the malware being analyzed. The model is generated using malware behavior observed by executing the malware in a controlled environment. The features embedded in the Markov chains are then extracted and used for classification. The goal is to minimize the number of states composing the Markov chains and consequently the size of the feature space. Action selection for triggering malware behavior is random.

The random action selection in these initial approaches limited their effectiveness in terms of execution path exploration since the action selection could not adapt to the malware behavior. More recent work in AMA proposes employing more intelligent methods like reinforcement learning to train autonomous agents for action selection. These take a game-theoretic approach and model AMA as stochastic or Bayesian games between the malware and the analyzer agent.

SECUR-AMA \cite{SARTEA2020103303} treats AMA as a stochastic game and proposes a Monte Carlo Tree Search based approach to dynamically generate the malware model at runtime. This approach does not rely on human experts for action selection. Malware authors often inject noise in their execution traces to defeat detection. SECUR-AMA remains vulnerable to this issue and proposes some ways to ignore the possible noise.

BAMA \cite{10.5555/3398761.3398901} models malware analysis as a Bayesian game. The analyzer uses prior knowledge about malware families to select triggering actions that reflect the current belief regarding which family the malware being analyzed belongs to. The aim is to lower the uncertainty about a malware sample’s type rather than its behavior.

\section{Model Design}
\label{sec:model}

This section describes the design choices made for the malware model and the Bayesian Model Combination implementations.

\subsection{Malware Model}

The action space $A$ is obtained by extracting the list of intents from the malware manifest file. As such, the action space can potentially change for each malware sample analyzed. The root node represents the main activity where each android application is initialized. Further nodes on the tree represent Android API calls made in response to trigger actions selected by the analyzer. The final result is a call graph obtained for each malware sample analyzed. Paths on the model graph are possible execution paths of the malware and show the probability of reaching a particular terminal state assuming a specific set of actions chosen by the analyzer.

\section{Experimental Setup and Analysis Process}
\label{sec:experiment}

The goal of the analyzer is to build the Android API call graph of the malware sample being analyzed with the minimal number of action selections required. The choice of the actions executed by the analyzer is based on the Bayesian Model Combination implementation running on a copy of the current malware model, which is updated at the end of every round. The action selected by the analyzer is executed on the running malware sample and the response, in the form of a sequence of API calls is stored. Each trace is represented as a path, starting at Init(), and is used to update the malware model graph and node statistics, including the transition probabilities between consecutive API calls. The analysis is stopped after a fixed number of trigger actions have been executed. The model thus generated is given as the final output. A linear SVM \cite{nissim2012detecting} is trained for classification on the call graphs obtained using different number of analyzer actions and the $F_1$ scores are used to determine the optimal number of analyzer actions required.

The experiments are conducted on an emulated Android 10 image instrumented using Frida \cite{Frida} to gather the API calls. The machine used to carry out the experiments is configured with an AMD Ryzen 5 3600 6-core processor at 3.6 GHz, 32 GB of RAM and an NVIDIA RTX 2070 Super with 8GB of GDDR6 memory. The analyzers are trained on a dataset containing 15000 malware samples collected from Androzoo \cite{Allix:2016:ACM:2901739.2903508}. The set contains 150 sample each from 100 different malware families.

\section{Evaluation}
\label{sec:evaluation}

The average time for the analyzer to select an action and observe the response was empirically observed to be about 11 seconds. After every observation, the emulated Android image is reset. In the experiments the time allotted to each malware sample is fixed at 5 minutes accounting for Android boot time, since a fresh virtual image is used for every new malware analysis, and the action selection and trace collection. We do not fix the size of the action space since a new list is extracted from each new sample. The analyzer thus considers a fresh action space for each malware and does not consider any other actions it might have encountered with other malware samples.

\begin{table}[htbp]
   \caption{Total Time taken per malware sample and optimal number of analyzer actions needed}
   \label{tab:uct_comp}
   \begin{center}
   \begin{tabular}{c c c}
     \toprule
     Analyzer Model & Analyzer Actions & Time [seconds] \\
     \midrule
     Bayes-UCT  & 8 & 168\\
     BMC-Constant $\epsilon$  & 8 & 168\\
     $\epsilon$-BMC  & 7 & 147\\
     \bottomrule
   \end{tabular}
   \end{center}
 \end{table}
 
  \begin{figure}[htbp]
    \centerline{\includegraphics[scale=0.6]{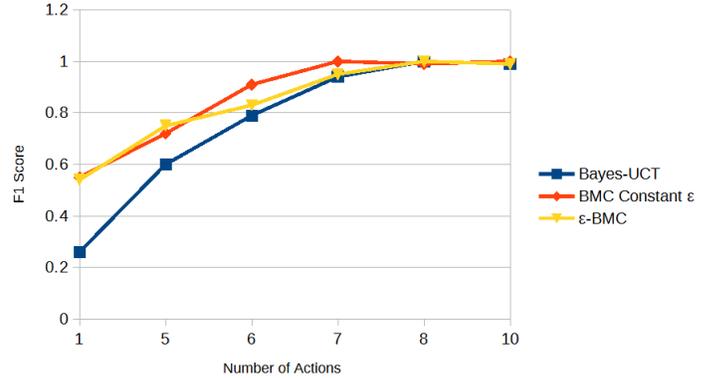}}
    \caption{F1-Scores for Linear SVM classifier}
    \label{f1score}
    \end{figure}
    
Table~\ref{tab:uct_comp} shows the optimal number of analyzer actions for BMC with a constant annealing approach to $\epsilon$, $\epsilon$-BMC and BayesUCT \cite{bayes_uct}. $\epsilon$-BMC gives an improvement compared to constant $\epsilon$ in terms of the number of actions required. BMC-Constant $\epsilon$ gives similar performance to BAMA \cite{10.5555/3398761.3398901} with the number of actions required being 8. The improvement offered by $\epsilon$-BMC, while small in terms of the number of actions required, becomes more significant when considering the time taken to generate the model when the number of actions is limited to the optimal number found, and scalled up to larger number of samples being analyzed. Fig.~\ref{f1score} shows the $F_1$-score values obtained by a linear SVM trained for classification on the call graphs generated and shows how performance is affected by the number of analyzer actions used to obtain the graphs. This allows us to choose the number of analyzer actions shown in Table \ref{tab:uct_comp}. The Bayesian Model combination approach described in this paper requires significantly lower number of analyzer actions to reach the best overall classification scores as compared to the stochastic game and UCT based approach of SECUR-AMA, even with different UCB algorithms \cite{uct_ama}.
 
\section{Conclusion}
\label{sec:conclusions}

Previous work in AMA has explored some Bayesian approaches by modeling AMA as a Bayesian game and using Bayesian learning for action selection in dynamic malware analysis. We consider another model for AMA and examine a Bayesian Model Combination based approach to the problem. $\epsilon$-BMC shows significant improvement over stochastic game models and some improvement over other Bayesian models considered in the literature. There is scope for improvements by considering Boltzmann exploration \cite{} and comparing against other adaptive annealing approaches.

\printbibliography

\end{document}